\documentclass[aps,prl,preprint,superscriptaddress]{revtex4-2}

\usepackage{multirow}
\usepackage{graphicx}
\usepackage{amsfonts,amsmath,amssymb,amsthm}
\usepackage{epstopdf}
\usepackage{upgreek,xspace}
\usepackage{chngcntr}
\usepackage{xr}
\usepackage{ulem}
\usepackage[colorlinks,allcolors=blue]{hyperref}
\usepackage[version=3]{mhchem} 

\begin{document}
\bibliographystyle{apsrev4-2}
\title{Quantum Hall effect in vacancy-engineered $\beta$-Ag$_2$Te}
\author{Mizuki~Ohno}
\affiliation{Department of Applied Physics and Materials Science, California Institute of Technology, Pasadena, California 91125, USA.}
\affiliation{Institute for Quantum Information and Matter, California Institute of Technology, Pasadena, California 91125, USA.}

\author{Veronica~Show}
\affiliation{Department of Applied Physics and Materials Science, California Institute of Technology, Pasadena, California 91125, USA.}
\affiliation{Institute for Quantum Information and Matter, California Institute of Technology, Pasadena, California 91125, USA.}

\author{Reiley~Dorrian}
\affiliation{Department of Applied Physics and Materials Science, California Institute of Technology, Pasadena, California 91125, USA.}
\affiliation{Institute for Quantum Information and Matter, California Institute of Technology, Pasadena, California 91125, USA.}

\author{Joseph~Falson}
\email{falson@caltech.edu}
\affiliation{Department of Applied Physics and Materials Science, California Institute of Technology, Pasadena, California 91125, USA.}
\affiliation{Institute for Quantum Information and Matter, California Institute of Technology, Pasadena, California 91125, USA.}

\begin{abstract}
Accessing surface quantum transport in topological insulators is hampered by residual bulk conduction arising from lattice defects. Here, we demonstrate a novel synthesis pathway for realizing high mobility $\beta$-Ag$_2$Te thin films where surface transport is dominant. An \textit{in-situ} vacancy engineering step as part of the molecular beam epitaxy growth process acts to modify the stoichiometry and suppress donor-type defects, enabling continuous tuning of the sheet carrier density over more than an order of magnitude through the charge-neutrality point without an external gate electrode. In the lower-carrier-density films, a fully developed $\nu=1$ quantum Hall state is observed, and Landau-level energies extracted across samples collapse onto the $E_N=v_\mathrm{F}\sqrt{2e\hbar NB}$ relation, providing evidence for the massless Dirac dispersion of the top and bottom surface states. These results establish stoichiometry-driven vacancy engineering as a versatile lithography- and gate-free approach to accessing quantum Hall transport in epitaxial topological-insulator thin films.
\end{abstract}

\maketitle
\noindent
\section*{Introduction}
The linearly dispersing Dirac surface states within the bulk band gap define topological insulators (TIs) as a unique phase of electronic matter~\cite{hasan2010, qi2011, Bansil2016_Colloquiuma}.
When the chemical potential lies within the gap, these surface states govern conductivity and, given their unique dispersion, can give rise to a half-integer-shifted quantum Hall effect (QHE) at sufficiently high magnetic fields~\cite{Xu2014_observation, Yoshimi2015, Koirala2015, Moon2018_Solution, Salehi2019_QuantumHall, Yoshimi2025_Emergence}. Accessing this surface-dominated regime experimentally is challenging due to the narrow band gap in most TIs. 
In the prototypical \ce{Bi2Se3} family, bbulk carrier densities commonly reach $\sim 10^{19}$~cm$^{-3}$ even in nominally undoped crystals~\cite{Hor2009_pdoping, Analytis2010_Twodimensional, ren2011} due to crystalline disorder, thereby masking features which are uniquely attributable to surface states. To suppress bulk conduction, previous studies have employed electrostatic gating~\cite{Chang2013_QHE, Xu2014_observation, Yoshimi2015}, compensation doping~\cite{ren2011, Ren2010_Optimizing, Zhang2011_BSTS, Kushwaha2016_SnBSTS}, and reduction of sample dimensions to the nanoscale~\cite{Qu2010_Quantum, Cho2011_Nanoribbon, Xiu2011_Nanoribbon}; yet each strategy introduces trade-offs---fabrication complexity and band bending at the gate-insulator interface, additional disorder, or limitations on scalability---that motivate the search for alternative approaches based on intrinsic material design.

\begin{figure}[h]
    \includegraphics[width=8.5cm]{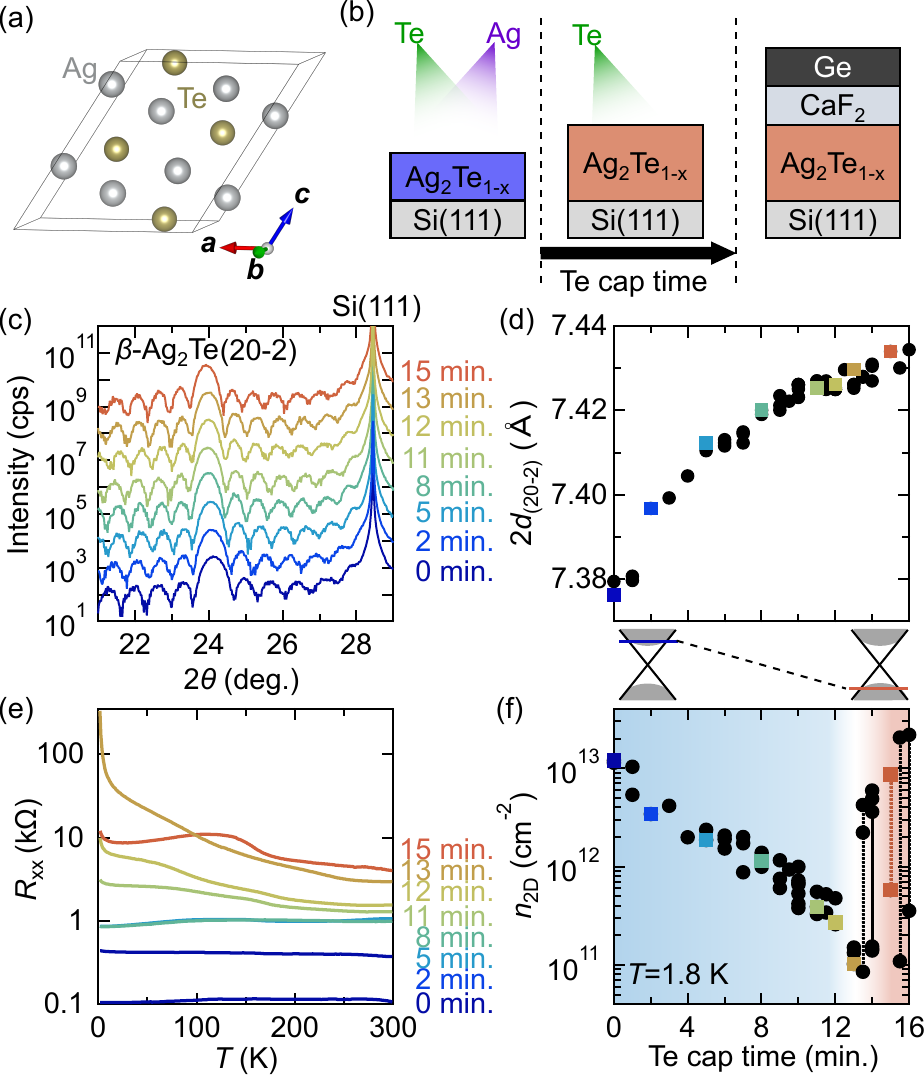}
    \caption{
        (a) Crystal structure of $\beta$-Ag$_2$Te drawn with VESTA~\cite{Momma2011}.
        (b) Schematic illustration of the growth process involving a coevaporation step (left), a Te-only deposition period (center), and a protective bilayer of CaF$_2$ and Ge (right).
        For differing Te-cap times,
        (c) X-ray diffraction $\theta$--$2\theta$ scans around the $\beta$-Ag$_2$Te ($20$-$2$) reflection,
        (d) ($20$-$2$) out-of-plane lattice spacing $2d_{(20\text{-}2)}$,
        (e) Temperature dependence of sheet resistance $R_\mathrm{xx}$, and
        (f) sheet carrier density $n_\mathrm{2D}$ measured at $T=1.8$~K, with blue and red shaded regions indicating the $n$-type and $p$-type dominated regimes.
        Tie-lines between data points in the $p$-type regime correspond to multicarrier fit results.
        The schematic above (f) illustrates the shift of the Fermi level from the conduction band ($n$-type) to the valence band ($p$-type) as the Te-cap time increases.
        Colored squares correspond to samples shown in (c).
    }
\label{fig:1}
\end{figure}

Among TIs, the surface Dirac cone of $\beta$-Ag$_2$Te is characterized by a highly anisotropic dispersion reflecting its low-symmetry monoclinic structure (Fig.~\ref{fig:1}(a)) and a small electron effective mass on the order of $0.01\,m_0$, where $m_0$ is the free-electron mass~\cite{Xu1997_Large, Schnyders2000_Magnetoresistanceb, Lee2002_BandGape, Hu2005_Current, Zhang2011_Topological, Lee2012_Single, Sulaev2013_Experimental, Sulaev2015_Gatetuneda, Leng2020_GateTunable}. Despite a relatively narrow bulk gap reported in the range of 64--100~meV depending on stoichiometry and measurement technique~\cite{Dalven1966_Energyb, Vassilev1999_Electrical, Zhang2011_Topological, Sulaev2013_Experimental, Leng2023_Nondegenerate, Ai2024}, $\beta$-Ag$_2$Te offers a distinct pathway to engineering low bulk carrier densities through native defect control. Specifically, the bulk carrier type is determined by the Ag/Te stoichiometry: Ag-rich samples exhibit $n$-type conduction driven by shallow donor-type Ag interstitials (\ce{Ag_i}), while Te-rich (Ag-deficient) samples exhibit $p$-type conduction governed by Ag vacancies (\ce{V_{Ag}})~\cite{Gottlieb1960_Electrical, Taylor1961_Thermoelectrica, Dalven1966_Energyd, Schnyders2000_Magnetoresistanceb, Lee2002_BandGape, Sun2003_Gianta, Aliev2015_Dependence, Leng2020_GateTunable, Hirata2023_Magnetothermala, Wuliji2023_Study}. Stoichiometric optimization, therefore, emerges as an intrinsic lever for accessing the surface-dominated quantum-transport regime without resorting to external gating or extrinsic dopants. Despite this conceptual advantage, prior quantum-transport studies have been restricted to exfoliated nanoflakes with thicknesses above ${\sim}70$~nm, where stoichiometry tuning is inherently challenging and residual bulk contributions prevent full quantization of the Hall effect~\cite{Leng2023_Nondegenerate}. Existing thin-film studies, in turn, have focused on structural and basic transport characterization without entering the quantum-transport regime~\cite{Dalven1966_Vacuuma, Das1983_Semiconductinga, Chuprakov1998_Largeb, Nakaya2020_Singlecrystalline, Takegawa2026_Epitaxial}. As a result, stoichiometry-engineered thin films have remained an unexplored platform for realizing surface-dominated transport in this material, and a continuous, reproducible route to tune the Fermi level across the bulk gap in epitaxial films has been lacking.

In this work, we overcome these limitations by demonstrating vacancy engineering of films via an \textit{in situ} Te capping step (Fig.~\ref{fig:1}(b)) using molecular beam epitaxy (MBE). By exploiting the high \ce{Ag^+}-ion diffusivity of $\beta$-Ag$_2$Te~\cite{Miyatani1958_Ionic, Hull2004_Superionics}, Te capping progressively reduces the concentration of donor-type Ag interstitials at room temperature. This enables continuous tuning of the sheet carrier density ($n_\mathrm{2D}$) through the charge-neutrality point without the use of an external gate electrode. All films we report are $\beta$-Ag$_2$Te films of 13--15~nm thickness---approximately half the thickness of the thinnest $\beta$-Ag$_2$Te sample previously measured electrically~\cite{Takegawa2026_Epitaxial}---while remaining thicker than the estimated surface-state penetration depth, ensuring that the top and bottom surface states remain effectively decoupled (Supplementary Note~S8). This approach culminates in the resolution of a dissipationless QHE at filling factor $\nu=1$ and a $\sqrt{BN}$ Landau-level (LL) energy spectrum characteristic of massless Dirac fermions.

\section*{Results}
\subsection*{Film growth and structural characterization}
Due to the dramatically higher vapor pressure of Te relative to Ag, thin films of $\beta$-Ag$_2$Te are not naturally amenable to an adsorption-controlled growth mode where Te is overdosed. We therefore begin the crystal growth under Te-deficient conditions on Si(111) substrates (see Methods for details). Following this initial deposition step, a Te layer is deposited \textit{in situ} at room temperature for a controlled duration. The Te layer nourishes the crystal and induces a systematic shift in X-ray diffraction (XRD) conditions. Figure~\ref{fig:1}(c) shows, in addition to sharp reflections from the $\beta$-Ag$_2$Te ($20$-$2$) planes without any impurity phases, a monotonic shift in the ($20$-$2$) peak towards lower angle. This corresponds to a dilation of the interplanar spacing $2d_{(20\bar{2})}$ (Fig.~\ref{fig:1}(d)), consistent with Te atoms shifting the stoichiometry away from Te-deficient conditions (Supplementary Note~S1). Finally, a bilayer of \ce{CaF2} (${\sim}\,10$~nm) and amorphous Ge (${\sim}\,5$~nm) is deposited \textit{in situ} for protection (Fig.~\ref{fig:1}(b)). The total film thickness ranges from 13--15~nm and increases with increasing Te-cap time.

The Te-dosing step also tunes the carrier concentration and type.
The temperature-dependent sheet resistance $R_\mathrm{xx}$ of the films is shown in Fig.~\ref{fig:1}(e), while the sheet carrier density $n_\mathrm{2D}$ obtained through Hall measurements is plotted as a function of Te-cap time in Fig.~\ref{fig:1}(f).
The rising resistance with increasing cap time is rationalized by the decline in carrier concentration from above $10^{13}$~cm$^{-2}$ towards $10^{11}$~cm$^{-2}$ until the cap time exceeds 13~min, where an increase in $n_\mathrm{2D}$ appears concomitant with a reduction in $R_\mathrm{xx}$ at low temperatures. An analysis of the sign of the Hall coefficient highlights that this indicates a crossover from $n$- to $p$-type dominated transport, consistent with a Fermi-level shift across the bulk gap illustrated schematically above Fig.~\ref{fig:1}(f).

The same stoichiometry--carrier-type correlation has previously been reported in bulk crystals, where the Ag/Te ratio is coarsely tuned between Ag-rich and Ag-deficient growth conditions~\cite{Gottlieb1960_Electrical, Taylor1961_Thermoelectrica, Dalven1966_Energyd, Schnyders2000_Magnetoresistanceb, Aliev2015_Dependence, Hirata2023_Magnetothermala, Wuliji2023_Study}, and in exfoliated nanoflakes, where $n$- and $p$-type flakes are obtained serendipitously from different growth-temperature regimes~\cite{Leng2020_GateTunable}. In contrast, the present approach tunes a single sample series continuously through the charge-neutrality point by varying only the \textit{in situ} Te-cap duration, converting stoichiometry control from a discrete growth-parameter choice into a fine-grained post-growth knob. The convergence of structural and transport trends provides evidence that Te capping systematically reduces the concentration of donor-type defects (Supplementary Note~S2).

\subsection*{Carrier-density tuning by vacancy engineering}

\begin{figure*}[h]
    \includegraphics[width=17cm]{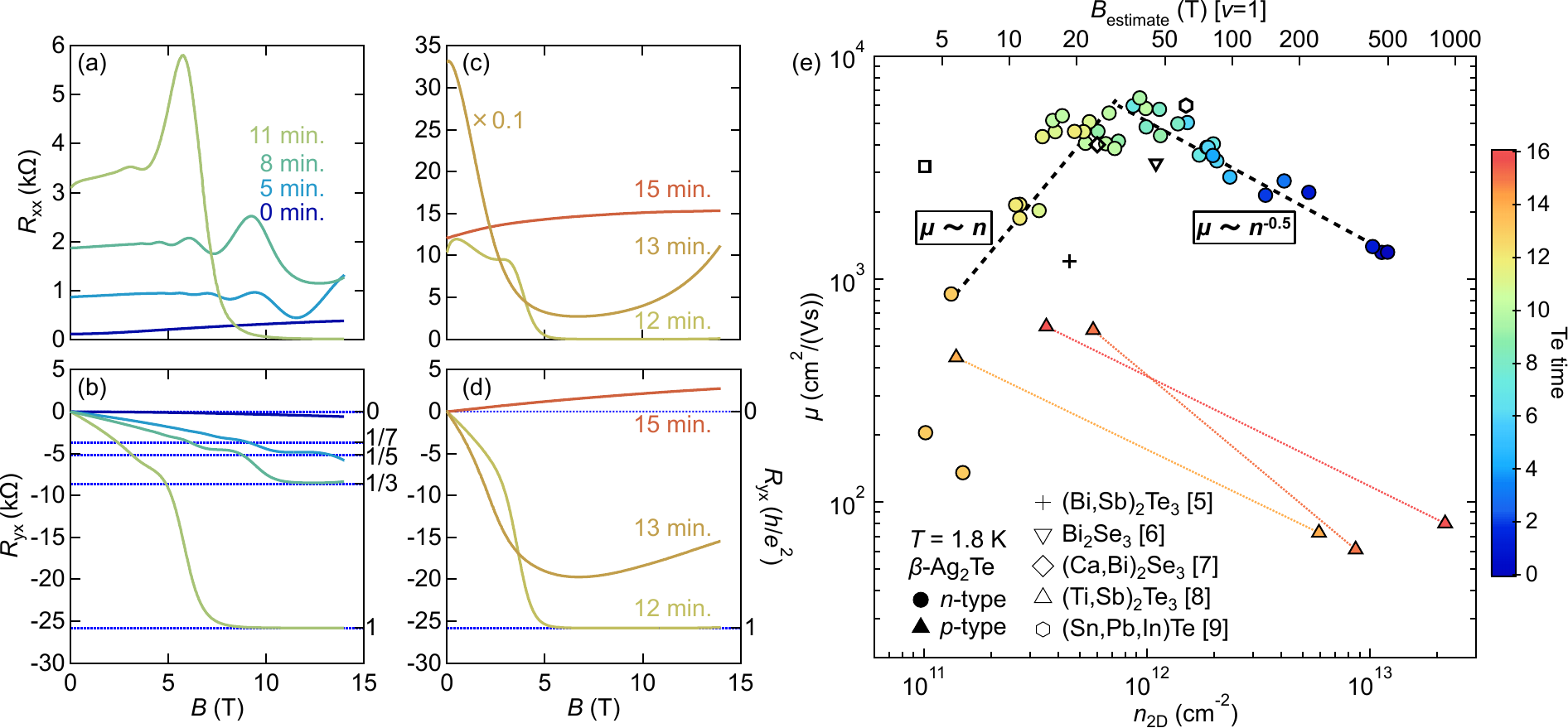}
    \caption{
        Out-of-plane $B$-field dependence of (a,c) longitudinal resistance $R_\mathrm{xx}$ and (b,d) Hall resistance $R_\mathrm{yx}$ with Te-cap times of 0, 5, 8, and 11~min (a,b) and 12, 13, and 15~min (c,d); the 13-min curve in (c) is scaled by 0.1 for clarity.
        Dotted horizontal lines in (b,d) indicate the quantized Hall resistances at filling factors $\nu=1$, $3$, and $5$.
        (e) Hall mobility $\mu$ versus two-dimensional carrier density $n_\mathrm{2D}$ for all samples, colored by Te-cap time.
        Circles and triangles indicate $n$-type and $p$-type samples, respectively.
        Solid lines indicate $\mu\propto n_\mathrm{2D}$ and $\mu\propto n_\mathrm{2D}^{-0.5}$ as guides to the eye.
        For the $p$-type samples, multicarrier fitting was performed, and the two carriers are connected by dashed lines; for visual clarity, only three representative samples are plotted.
        Other open markers show literature data for \ce{Bi2Se3}-family TI systems~\cite{Salehi2019_QuantumHall, Koirala2015, Moon2018_Solution, Yoshimi2015, Yoshimi2025_Emergence}.
    }
\label{fig:2}
\end{figure*}

Figure~\ref{fig:2} summarizes the magnetotransport properties of representative samples spanning the full range of Te-cap times, recorded at $T=1.8$~K. Figures~\ref{fig:2}(a)--(d) present longitudinal resistance $R_\mathrm{xx}$ and Hall resistance $R_\mathrm{yx}(B)$ and illustrate that, as the cap time increases from 0 to 11~min (Figs.~\ref{fig:2}(a),(b)), $R_\mathrm{yx}(B)$ evolves from a nonlinear, multi-carrier response into a strictly linear field dependence, reflecting the progressive suppression of bulk conduction and the emergence of single-carrier surface-state transport (Supplementary Note~S3). For samples with cap times between 5 and 11~min, Shubnikov--de Haas (SdH) oscillations with quantized Hall responses emerge, and in the 11--12~min window a fully developed $\nu=1$ quantum Hall plateau is resolved at $R_\mathrm{yx}=h/e^2$ with vanishing $R_\mathrm{xx}$ (Figs.~\ref{fig:2}(c),(d)).
At a cap time of 13~min the $\nu=1$ quantum Hall plateau is markedly weakened, with $R_\mathrm{xx}$ no longer vanishing and $R_\mathrm{yx}$ deviating from $h/e^2$, and any indications of the QHE are completely absent by 15~min capping time, along with a change in sign of the Hall resistance.

The Hall mobility $\mu$ as a function of $n_\mathrm{2D}$ is plotted in Fig.~\ref{fig:2}(e), revealing three distinct transport regimes: high-mobility $n$-type samples displaying quantum Hall features, samples in which carrier localization is prominent, and $p$-type samples where the chemical potential enters the valence band. For the high-mobility $n$-type samples, two power-law branches are separated by a mobility maximum at $n_\mathrm{2D}\approx 1\times 10^{12}$~cm$^{-2}$. In the high-density regime ($n_\mathrm{2D}\gtrsim 1\times 10^{12}$~cm$^{-2}$), $\mu$ follows $\mu\propto n_\mathrm{2D}^{-0.5}$, the exponent expected for Dirac carriers when momentum relaxation is dominated by short-range disorder under strongly screened long-range Coulomb potentials~\cite{Hwang2007_Carrierb, DasSarma2011}. Below $n_\mathrm{2D}\sim 1\times 10^{12}$~cm$^{-2}$, the trend reverses to $\mu\propto n_\mathrm{2D}$, the characteristic signature of the charge-puddle regime near the Dirac point, where the vanishing density of states suppresses Thomas--Fermi screening and the system fragments into electron--hole puddles~\cite{Hwang2007_Carrierb, DasSarma2011, Beidenkopf2011, Skinner2012}. A quantitative treatment of both branches is provided in Supplementary Note~S3.
A distinctive feature of $\beta$-Ag$_2$Te is that interstitial silver \ce{Ag_i} simultaneously acts as the dominant donor and as the dominant charged-impurity scattering center, so Te capping tunes the carrier density and the impurity density in concert (Supplementary Note~S2); this self-regulating mechanism underlies the wide tunability and high mobility achieved in our films. The $p$-type samples exhibit mobilities approximately one order of magnitude lower than those of $n$-type samples at comparable carrier densities and show no resolvable quantum oscillations, consistent with the substantially heavier effective mass of the valence band relative to the conduction band~\cite{Aliev2002_Electrical, Aliev2015_Dependence, Leng2020_GateTunable}.

\subsection*{Analysis of quantum transport}

\begin{figure}[h]
    \includegraphics[width=8.5cm]{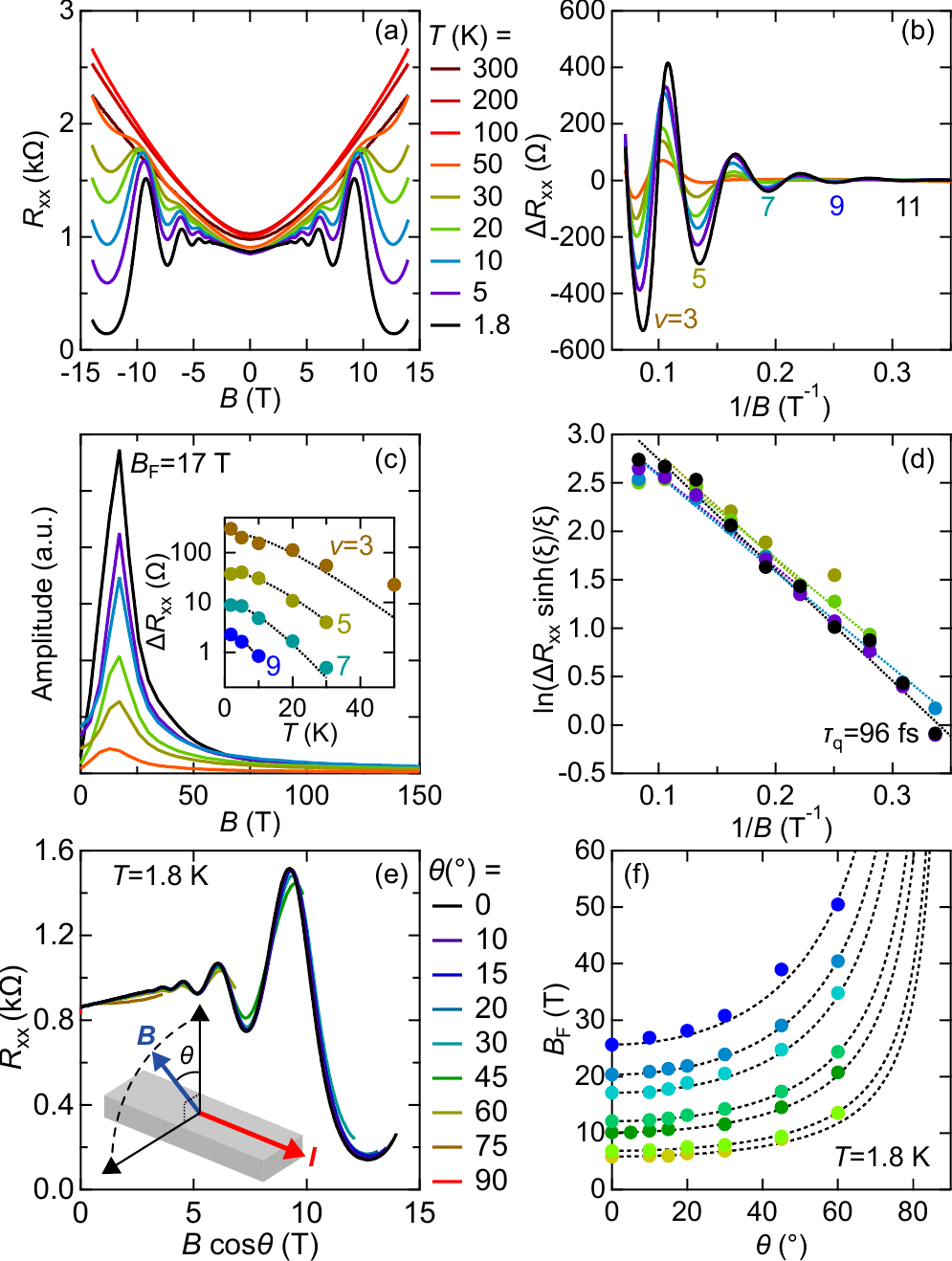}
    \caption{
        (a) $R_\mathrm{xx}$ versus $B$ from 1.8 to 300~K, and (b) the oscillatory component $\Delta R_\mathrm{xx}$ versus $1/B$ for $T\leq 50$~K with filling factors $\nu$ indicated.
        (c) Fast Fourier transform of (b), showing a single dominant Shubnikov--de Haas frequency $B_\mathrm{F}=17$~T; the inset shows the temperature dependence of $\Delta R_\mathrm{xx}$ at $\nu=3,\,5,\,7,\,9$, from which the cyclotron effective mass $m_\mathrm{eff}=0.061\,m_0$ is extracted by Lifshitz--Kosevich fitting (dotted lines).
        (d) Dingle plot of $\ln[\Delta R_\mathrm{xx}\sinh(\xi)/\xi]$ versus $1/B$ for $T\leq 30$~K, yielding the quantum scattering time $\tau_q=96$~fs from linear fits (dotted lines).
        (e) $R_\mathrm{xx}$ versus $B\cos\theta$ at 1.8~K for tilt angles $\theta=0$--$90^\circ$ (inset: geometry), and (f) $\theta$ dependence of $B_\mathrm{F}$ at 1.8~K for multiple samples, with the expected $1/\cos\theta$ behavior shown as dashed lines.
    }
\label{fig:3}
\end{figure}

We now focus on the 8-min Te-cap sample shown in Fig.~\ref{fig:3} for a representative analysis of the quantum transport. Figure~\ref{fig:3}(a) shows $R_\mathrm{xx}(B)$ at a series of temperatures with pronounced SdH oscillations clearly visible up to 50~K as a result of large LL gaps associated with the light Dirac carriers. The oscillatory component $\Delta R_\mathrm{xx}$ exhibits well-defined periodicity in $1/B$ (Fig.~\ref{fig:3}(b)), and the fast Fourier transform reveals a single dominant SdH frequency $B_\mathrm{F}=17$~T (Fig.~\ref{fig:3}(c)), indicating that transport is dominated by a single carrier pocket. The temperature dependence of the SdH amplitude at $\nu=3,\,5,\,7,\,9$ is well reproduced by the Lifshitz--Kosevich thermal damping factor (inset of Fig.~\ref{fig:3}(c)), yielding a cyclotron effective mass $m_\mathrm{eff}=0.061\,m_0$. From the Onsager relation $B_\mathrm{F}=\hbar k_\mathrm{F}^2/(2e)$, where $k_\mathrm{F}$ is the Fermi wave vector, and a linear dispersion, we obtain a Fermi velocity $v_\mathrm{F}\approx4.0\times10^5$~m\,s$^{-1}$ and a Fermi energy $E-E_\mathrm{DP}\approx65$~meV above the Dirac point energy $E_\mathrm{DP}$.
Dingle analysis (Fig.~\ref{fig:3}(d)) yields a quantum scattering time $\tau_q=96$~fs, which characterizes the total scattering rate without angular weighting; the ratio $\mu/\mu_q\sim1$--$2$ across all samples (where $\mu_q=e\tau_q/m_\mathrm{eff}$ is the quantum mobility) is consistent with relatively short-range dominant disorder (Supplementary Note~S9).

To verify the dimensionality of the carriers, angle-dependent magnetotransport measurements were performed by rotating the magnetic field from the out-of-plane ($\theta=0^\circ$) toward the in-plane direction ($\theta=90^\circ$), while keeping $I\perp B$ throughout (inset of Fig.~\ref{fig:3}(e)). When plotted as a function of $B\cos\theta$, all $R_\mathrm{xx}$ oscillation curves collapse onto a single trace (Fig.~\ref{fig:3}(e))---the definitive signature of a two-dimensional Fermi surface. The extracted $B_\mathrm{F}$ follows the $1/\cos\theta$ dependence consistently across all measured samples (Fig.~\ref{fig:3}(f)), confirming the two-dimensional nature of the conduction (Supplementary Note~S6).

\begin{figure}[h]
    \includegraphics[width=8.5cm]{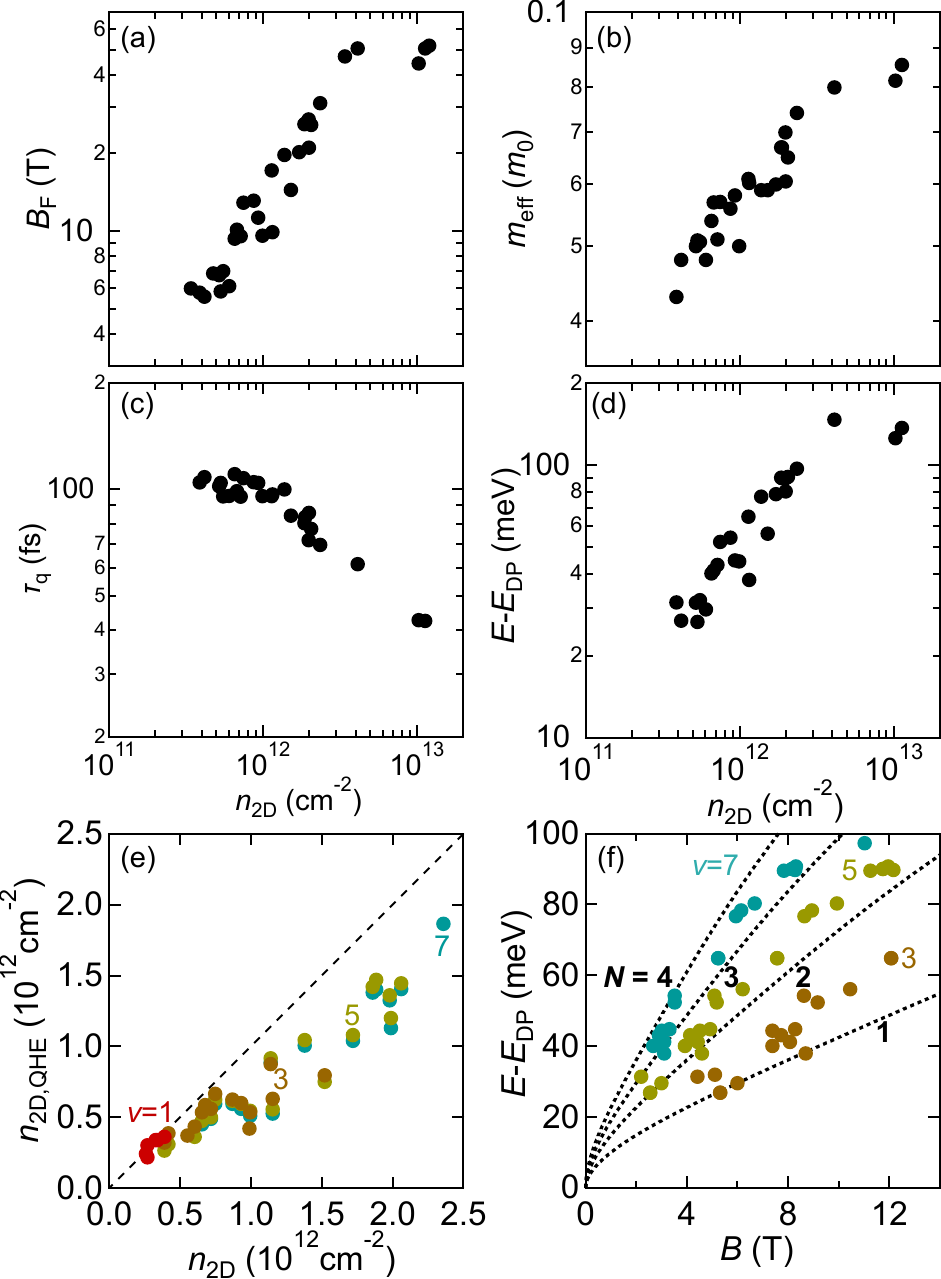}
    \caption{
        Topological surface-state parameters and the quantum Hall effect at 1.8~K.
        $n_\mathrm{2D}$ dependence of (a) $B_\mathrm{F}$, (b) $m_\mathrm{eff}$, (c) $\tau_q$, and (d) Fermi energy $E-E_\mathrm{DP}$ for all measured samples.
        (e) Carrier densities from the low-field Hall slope ($n_\mathrm{2D}$) versus those from the $\nu=1,\,3,\,5,\,7$ quantum Hall states ($n_\mathrm{2D,\,QHE}$); the dotted line denotes $n_\mathrm{2D,\,QHE}=n_\mathrm{2D}$.
        (f) Fermi energy $E-E_\mathrm{DP}$ versus the magnetic field of the $\nu=1,\,3,\,5,\,7$ QHE plateaus, with dotted lines showing the LL energies $E_N=v_\mathrm{F}\sqrt{2e\hbar NB}$ ($N=1,2,3,4$) expected for a linear dispersion.
    }
\label{fig:4}
\end{figure}

We now interrogate an extended set of samples in order to amass evidence for the Dirac nature of carriers. Figures~\ref{fig:4}(a)--(d) summarize the carrier-density dependence of $B_\mathrm{F}$, $m_\mathrm{eff}$, $\tau_q$, and $E-E_\mathrm{DP}$. $B_\mathrm{F}$ scales linearly with $n_\mathrm{2D}$, as expected for a single two-dimensional Fermi pocket. Both $m_\mathrm{eff}$ and $E-E_\mathrm{DP}$ follow a $\sqrt{n_\mathrm{2D}}$ dependence, the defining signature of a linear (non-parabolic) Dirac dispersion, for which $m_\mathrm{eff}=\hbar k_\mathrm{F}/v_\mathrm{F}\propto\sqrt{n_\mathrm{2D}}$ and $E-E_\mathrm{DP}=\hbar v_\mathrm{F}k_\mathrm{F}\propto\sqrt{n_\mathrm{2D}}$ through $k_\mathrm{F}^{2}\propto n_\mathrm{2D}$. $\tau_q$ increases monotonically with decreasing $n_\mathrm{2D}$, consistent with the progressive suppression of small-angle scattering as the \ce{Ag_i} density is reduced by Te capping.

First-principles calculations place the size of the inverted band gap in $\beta$-Ag$_2$Te at $E_\mathrm{g}\approx80$~meV~\cite{Zhang2011_Topological, Sulaev2013_Experimental, Leng2023_Nondegenerate, Ai2024}.
Our data show that samples with $n_\mathrm{2D}\lesssim6\times10^{11}$~cm$^{-2}$, corresponding to $E-E_\mathrm{DP}\lesssim40$~meV, exhibit $R_\mathrm{xx}$ that approaches zero at high field.
This threshold coincides with the Fermi level being located within $\sim E_\mathrm{g}/2$ of the middle of the bulk gap, indicating that bulk conduction is effectively suppressed in this regime---precisely the condition under which the $\nu=1$ quantum Hall plateau identified above becomes resolvable.

We further examine the role of surface versus bulk conduction in Fig.~\ref{fig:4}(e) by comparing $n_\mathrm{2D}$ extracted from the Hall analysis with that obtained from the magnetic fields at which quantum Hall plateaus appear at $\nu=1,3,5,7$ ($n_\mathrm{2D,\,QHE}$).
The former will contain both surface and bulk contributions, while the latter exclusively reflects high-mobility carriers contributing to the quantum Hall physics.
In the lowest-density samples, $n_\mathrm{2D,\,QHE}$ extracted from the $\nu=1$ plateau agrees with $n_\mathrm{2D}$ (dotted line), confirming that a single set of surface carriers governs both the low-field and the quantized transport and that residual bulk conduction is negligible.
In higher-density samples however, $n_\mathrm{2D,\,QHE}$ obtained from $\nu=3,5,7$ falls systematically below $n_\mathrm{2D}$, consistent with a residual bulk-parallel channel that enhances the low-field Hall slope without contributing to the quantized surface transport (Supplementary Note~S7).

In contrast to parabolic or conventional Dirac systems such as graphene, where a spin-degenerate Fermi sea develops LLs split by a Zeeman term, a TI with two decoupled surfaces experiences contributions to the Hall conductivity from each surface according to $\sigma_{xy}^{(\mathrm{single})}=(N+1/2)\,e^2/h$, with the $N$-th LL of a massless Dirac fermion at
\begin{equation}
  E_N = v_\mathrm{F}\sqrt{2e\hbar|N|B}.
  \label{eq:LL}
\end{equation}
This results in a series where $\nu=1,\,3,\,5,\,7,\ldots$ elicit the largest gaps.
Figure~\ref{fig:4}(f) plots $E-E_\mathrm{DP}$ extracted from SdH analysis against the magnetic field at which the QHE plateaus are observed for all measured samples.
The data from all samples collapse onto the Dirac LL curves (Eq.~\ref{eq:LL}) for $N=1,2,3,4$ (dotted lines), demonstrating the $\sqrt{BN}$ scaling characteristic of massless Dirac fermions---in sharp contrast to the linear-in-$N$ spectrum of conventional two-dimensional electron gases.
The observed plateaus $\nu=3,5,7$ fall between $N=1$ and $2$, $N=2$ and $3$, and $N=3$ and $4$, in quantitative agreement with the two-surface relation $\nu=(N_\mathrm{top}+1/2)+(N_\mathrm{bottom}+1/2)$.
The strict odd-integer sequence, together with the single SdH frequency $B_\mathrm{F}$ (Fig.~\ref{fig:3}(c)) and the agreement between $n_\mathrm{2D}$ and $n_\mathrm{2D,\,QHE}$ in the lowest-density samples, thus provides consistent evidence that the top and bottom surfaces share nearly identical Fermi energies, made physically plausible by the near-symmetric dielectric environment provided by the Si(111) substrate and the Ge/\ce{CaF2} capping (Supplementary Note~S12).

\section*{Discussion}
Compared with semiconductor quantum wells, TIs tend to be relatively disordered, which is reflected in their modest electron mobility.
Despite this, the light carriers from Dirac surface states host large Landau gaps, allowing quantum transport to be observed up to relatively high temperatures, as evidenced in Fig.~\ref{fig:3}, where distinct oscillations are apparent up to $T$=50~K in our $\beta$-Ag$_2$Te films.
Figure~\ref{fig:2}(e) compares $\mu$ versus $n_\mathrm{2D}$ for prominent TIs.
In \ce{Bi2Se3}-family TI systems, accessing the low-$n_\mathrm{2D}$, high-$\mu$ regime suitable for quantum Hall transport has typically required combinations of dedicated underlayers and capping layers, compensation doping with Ca or Ti, and electrostatic gating to fine-tune the chemical potential~\cite{Salehi2019_QuantumHall, Koirala2015, Moon2018_Solution, Yoshimi2015, Yoshimi2025_Emergence}.
In contrast, our $\beta$-Ag$_2$Te films reach a comparable regime using only an \textit{in situ} Te-capping layer, without lithographic patterning or gate tuning.
This simplification is rooted in the self-regulating mechanism unique to $\beta$-Ag$_2$Te (Supplementary Note~S2): because \ce{Ag_i} simultaneously serves as the dominant donor and the dominant scattering center, a single Te-capping step reduces both $n_\mathrm{2D}$ and the charged-impurity scattering rate in concert, exploiting the superionic \ce{Ag^+} diffusivity of $\beta$-Ag$_2$Te~\cite{Miyatani1958_Ionic, Hull2004_Superionics}.
$\beta$-Ag$_2$Te therefore emerges as a simple and chemically self-tuned platform for topological quantum transport, achieving a ($\mu$, $n_\mathrm{2D}$) balance competitive with the most heavily optimized TI films.

The complete absence of even-integer plateaus and the strict odd-integer sequence $\nu=1,3,5,7$ indicates that the two Dirac surfaces are effectively decoupled at the $\sim 14$~nm film thickness.
This is consistent with a surface-state penetration depth $\lambda\approx\hbar v_\mathrm{F}/E_\mathrm{g}\approx 2.5$--$4.5$~nm well below the film thickness~\cite{Linder2009_Anomalous}, and with the near-symmetric dielectric environment of the Si(111)/\ce{CaF2} heterostructure that maintains near-degeneracy of the two surfaces (Supplementary Notes~S8 and S12).
The combination of effective decoupling and energy degeneracy selects the half-integer-summed odd sequence $\nu=N_\mathrm{top}+N_\mathrm{bottom}+1$ over an independently quantized integer sequence, and such a simultaneous realization is rarely achieved in non-magnetic TI thin films.

\section*{Conclusion}
We have observed the dissipationless QHE in $\beta$-Ag$_2$Te attributable to Dirac surface states by developing a vacancy-engineering approach for thin films.
Despite the relatively small bulk gap of $\beta$-Ag$_2$Te, our gate- and lithography-free approach allows the chemical potential to be tuned with precise control across bulk- and surface-dominated transport regimes.
This work establishes a new material platform for integrating the topologically protected electronic modes of TIs with additional circuit elements.


\section*{Methods}

\subsection*{Film growth}

$\beta$-Ag$_2$Te thin films were synthesized in a MBE system with a base pressure of ${\sim}10^{-10}$~mbar on Si(111) substrates that had been annealed at $1000\,^{\circ}\mathrm{C}$ prior to growth to remove the native oxide.
The Si surface was subsequently flushed with Te to saturate dangling bonds and promote van der Waals epitaxy~\cite{Koma1992_Van, Koma1999_Van, Kampmeier2015_Suppressing}.
Te was supplied from a thermal cracker cell (tank temperature ${\sim}\,380\,^{\circ}\mathrm{C}$, cracking-zone temperature $1000\,^{\circ}\mathrm{C}$), and Ag was evaporated from a conventional effusion cell at ${\sim}\,860\,^{\circ}\mathrm{C}$.
The growth temperature and the Te and Ag fluxes, monitored by an ionization gauge, were tuned to obtain single-phase films; the Ag/Te flux ratio was set to approximately $2$ to establish Te-deficient conditions and suppress Te-rich secondary phases.
The substrate temperature during growth was maintained at $120\,^{\circ}\mathrm{C}$.
The as-deposited $\beta$-Ag$_2$Te thickness was $\approx 13$~nm, corresponding to a growth rate of $\approx 0.04$~\AA/s.
After growth, a Te capping layer was deposited \textit{in situ} at room temperature for a controlled duration (0--16~min) to compensate donor-type defects and tune the carrier density; this step also increased the total $\beta$-Ag$_2$Te thickness to up to $15$~nm depending on the cap time.
To protect the films from degradation in air, all samples were subsequently capped \textit{in situ} with \ce{CaF2} (${\sim}\,10$~nm) and amorphous Ge (${\sim}\,5$~nm) layers deposited at room temperature from conventional Knudsen effusion cells.

\subsection*{Structural and transport characterization}

Structural properties of the films were characterized by XRD (SmartLab, Rigaku) with Cu~K$\alpha$ radiation in a $\theta$--$2\theta$ configuration.
Electrical transport measurements above 1.8~K were carried out in a Quantum Design DynaCool Physical Properties Measurement System (PPMS) equipped with a 14~T superconducting magnet, using the standard resistivity option.
Each film was cut into a rectangular Hall-bar piece ($2\times5$~mm) and contacted with Al wires at six points via indium paste for simultaneous four-terminal measurements of $R_\mathrm{xx}$ and $R_\mathrm{yx}$.
For standard magnetotransport, the magnetic field was applied perpendicular to the film plane; the longitudinal and Hall components were separated by symmetrization and antisymmetrization with respect to the field direction.
For angle-dependent magnetotransport measurements, the magnetic field was rotated from the out-of-plane direction ($\theta=0^\circ$) to the in-plane direction ($\theta=90^\circ$) using a sample rotator, while keeping $I\perp B$.
Filling factors $\nu$ were assigned from the positions of the SdH minima in $R_\mathrm{xx}(B)$.
The cyclotron effective mass and Dingle temperature were determined from Lifshitz--Kosevich and Dingle analyses, as described in the main text.
For the LL energy-spectrum analysis (Fig.~\ref{fig:4}(f)), $E-E_\mathrm{DP}$ of each sample was obtained from $E-E_\mathrm{DP}=\hbar v_\mathrm{F}k_\mathrm{F}$, using the sample-specific $v_\mathrm{F}=\hbar k_\mathrm{F}/m_\mathrm{eff}$ and $k_\mathrm{F}=\sqrt{2eB_\mathrm{F}/\hbar}$ derived from $B_\mathrm{F}$ and $m_\mathrm{eff}$.


\section*{Acknowledgments}
We acknowledge funding provided by the Institute for Quantum Information and Matter, a NSF Physics Frontiers Center (NSF Grant PHY-2317110), and the Gordon and Betty Moore Foundation's EPiQS Initiative (Grant number GBMF10638).
We acknowledge the support of JSPS Overseas Research Fellowships.
This material is based upon work supported by the National Science Foundation Graduate Research Fellowship Program under Grant No. 2139433 (RD, VS).
Any opinions, findings, and conclusions or recommendations expressed in this material are those of the author(s) and do not necessarily reflect the views of the National Science Foundation.

\section*{Conflict of Interest}
The authors declare no conflict of interest.

\section*{References}
\bibliography{bib}

\end{document}